\documentclass[
pdflatex,
sn-mathphys-num,
referee,
]{sn-jnl}

\usepackage{graphicx}
\usepackage[version=4]{mhchem}
\usepackage{amsmath,amssymb,amsfonts}
\usepackage[normalem]{ulem}
\usepackage{soul}
\newcommand{\stkout}[1]{\ifmmode\text{\sout{\ensuremath{#1}}}\else\sout{#1}\fi}
\newcommand{\add}[1]{#1}
\newcommand{\del}[1]{}
\newcommand{\change}[2]{\del{#1}\add{#2}}

\newcommand{\uu}[1]{\,\ensuremath{\text{#1}}}
\usepackage{xspace}
\newcommand{\SI}{\textcolor{red}{Supplementary Information}\xspace}
\DeclareMathOperator\sgn{sgn}
\begin{document}

\title{\emph{Ab initio} modeling of TWIP and TRIP effects in $\beta$-Ti alloys}

\author*[1]{\fnm{David} \sur{Holec}}\email{david.holec@unileoben.ac.at}
\author[1]{\fnm{Johann} \sur{Grillitsch}}
\author[2]{\fnm{José L.} \sur{Neves}}
\author[1]{\fnm{David} \sur{Obersteiner}}
\author*[2]{\fnm{Thomas} \sur{Klein}}\email{thomas.klein@ait.ac.at}

\affil[1]{\orgdiv{Department of Materials Science}, \orgname{Montanuniversität Leoben}, \orgaddress{\street{Franz-Josef-Strasse 18}, \city{Leoben}, \postcode{8700}, \country{Austria}}}

\affil[2]{\orgdiv{LKR Light Metals Technologies}, \orgname{AIT Austrian Institute of Technology}, \orgaddress{\street{Lamprechtshausener Strasse 61}, \city{Ranshofen}, \postcode{5282}, \country{Austria}}}

\abstract{
Transformations in bcc-$\beta$, hcp-$\alpha$, and the $\omega$ phases of Ti alloys are studied using Density Functional Theory for pure Ti and Ti alloyed with Al, Si, V, Cr, Fe, Cu, Nb, Mo, and Sn. 
The $\beta$-stabilization caused by alloying Si, Fe, Cr, and Mo was observed, but the most stable phase appears between the $\beta$ and the $\alpha$ phases, corresponding to the martensitic $\alpha''$ phase. 
Next, the $\{112\}\langle11\bar1\rangle$ bcc twins are separated by a positive barrier, which further increases by alloying w.r.t. pure Ti. 
The $\{332\}\langle11\bar3\rangle$ twinning yields negative barriers for all species but Mo and Fe. 
This is because the transition state is structurally similar to the $\alpha$ phase, which is preferred over the $\beta$ phase for the majority of alloying elements. Lastly, the impact of alloying on twin boundary energies is discussed. These results may serve as design guidelines for novel Ti-based alloys with specific application areas.
}




\maketitle

\section{Introduction}
\label{sec:intro}
Titanium alloys generate a lot of interest due to a favorable combination of mechanical and structural properties, making them applicable in the aerospace industry~\cite{Najafizadeh2024-sb}.
Furthermore, their general biocompatibility also enables medical usage~\cite{Sidambe2014-kn}.
Recently, even stronger application-driven interest has been generated due to the additive manufacturing (AM) of parts with complicated geometries, as exemplified by the wire and arc additive manufacturing process~\cite{Lin2021-ic}.

\begin{figure}[hbt]
    \centering
    \includegraphics[width=\textwidth]{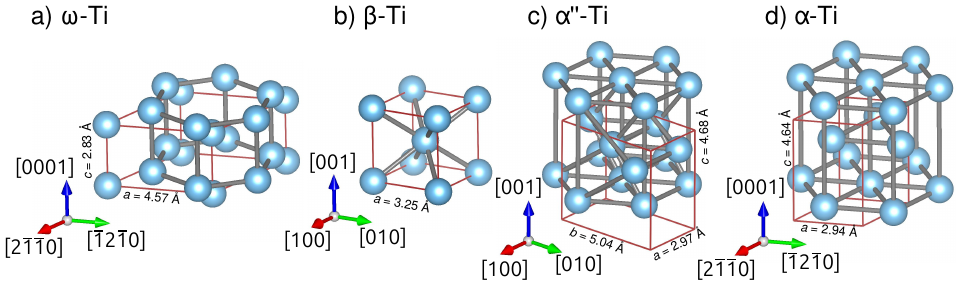}
    \caption{Unit cells of common Ti phases considered in this paper: a) $\omega$-Ti, b) bcc-$\beta$-Ti, c) martensitic $\alpha''$-Ti, and d) hcp-$\alpha$-Ti.
    The red lines denote the conventional cells used for their description.
    Visualized using VESTA~\cite{Momma2011-jk}.
    }
    \label{fig:unitcells}
\end{figure}

Titanium is an allotropic material~\cite{Leyens2006-ra}.
The two phases commonly associated with Ti are hcp-$\alpha$-Ti (space group: $P6_3/mmc$, Fig.~\ref{fig:unitcells}d) and bcc-$\beta$-Ti ($Im\bar3m$, Fig.~\ref{fig:unitcells}b), with the low-temperature $\alpha$-Ti maintaining stability up to $\approx 1155\uu{K}$~\cite{Dinsdale1991-uw}. 
The $0\uu{K}$ ground state structure is $\omega$-Ti ($P6/mmm$, Fig.~\ref{fig:unitcells}a), and its stability has been predicted to extend only to $\approx 186\uu{K}$ at ambient pressure~\cite{Mei2009-cx}.
Lastly, martensitic transformation upon rapid cooling from the $\beta$ phase field may yield either $\alpha''$ or $\alpha'$, depending on the amount of alloyed $\beta$-stabilizing elements in the Ti matrix.
The $\alpha'$ is an only slightly distorted $\alpha$ phase with lattice parameters close to the ideal hcp structure.
On the contrary, the $\alpha''$ ($Cmcm$, Fig.~\ref{fig:unitcells}c) resembles the lattice of the $\beta$ phase described in an orthorhombic frame, with atoms displaced (shuffled) away from their bcc positions towards the hcp arrangement~\cite{Zheng2022-iu}.

The mechanical response of Ti-based alloys is governed, next to dislocation motion, by triggering specific deformation mechanisms, namely Transformation-Induced Plasticity (TRIP) and Twinning-Induced Plasticity (TWIP)~\cite{Zhao2021-an}.
TRIP is a phase transformation upon applied stress from a metastable phase ($\beta$) to a martensitic one ($\alpha'$/$\alpha''$), thereby enhancing strength and ductility.
TWIP improves ductility by absorbing some of the applied strain in the form of deformation twins.
Although TWIP is usually associated with the stable $\alpha$ phase of Ti, it also can occur in the $\beta$ phase.
The prominent twinning mechanisms involve the $\{112\}\langle11\bar1\rangle$ and $\{332\}\langle11\bar3\rangle$ systems~\cite{Kwasniak2022-xv}, where the latter has been proposed to happen more likely due to a lower extent of required shuffling~\cite{Tobe2014-gh}.

Which deformation mechanism dominates is often assessed using the semi-empirical $Bo-Md$ method~\cite{Abdel-Hady2006-mj, Ren2018-yt, Lilensten2020-nz, Zhao2021-an}.
\add{
The $Bo-Md$ model was first suggested by \citet{Morinaga1988-wm}.
It has been shown to be effective in achieving the desired dominant deformation mechanism in several newly designed titanium alloys~\cite{Lilensten2020-nz, Marteleur2012-ti}. 
The method uses two calculated parameters.
$Bo$ is the bond order measure of the covalent bond strength between the Ti matrix and the alloying elements.
$Md$ is the correlation of the electronegativity of the metallic radii of the elements.
Both of these parameters have been calculated and tabulated for each element.
These parameters are then used to plot the alloys in the form of maps (see e.g.~\cite{Lilensten2020-nz, Marteleur2012-ti}). 
The original work by \citet{Morinaga1988-wm} also evidences how the position of an alloy is changed in this map by additions of a chemical element (see Fig. 9 in their work). 
This method offers a straightforward starting point for screening alloy compositions, particularly when targeting TRIP or TWIP behavior in metastable $\beta$-titanium alloys. 
For example, \citet{Lilensten2020-nz} successfully designed a dual-phase Ti-7Cr-1.5Sn alloy that exhibits both TRIP and TWIP mechanisms. 
Their alloy design approach combined the $Bo–Md$ method with CALPHAD-based thermodynamic modeling. 
Experimental validation of the alloy's deformation behavior, including stress-induced $\alpha''$ martensite, $\{33\bar2\}\langle11\bar3\rangle$ twinning, and at larger strains, $\{11\bar2\}\langle11\bar1\rangle$ twinning and $\omega$ phase formation, was achieved using in situ synchrotron X-ray diffraction and EBSD analysis. 
Applying the $Bo-Md$ method to elements considered in the present study, it can be concluded that the combined use of elements from both groups (V, Fe, Cr, Nb, Mo) and (Cu, Al, Sn) is necessary to induce twinning instead of conventional dislocation plasticity.
}

\add{
However, recent experimental investigations have increasingly highlighted the limitations of purely empirical models. 
For instance, \citet{Ballor2023-ti} highlight in their comprehensive review multiple cases where alloys with similar Mo-equivalent values exhibited significantly different deformation responses, ranging from $\alpha''$ formation to unexpected $\omega$ phase precipitation. 
Such discrepancies emphasize the fact that models like Mo-equivalency and $Bo-Md$, while valuable, do not fully capture the complexity of phase transformations and deformation behavior. 
}

\change{However, a}{A} recently published machine-learning-based approach has been claimed to supersede its prediction accuracy~\cite{Coffigniez2024-lm}.
Another indicator for the transformation mechanism is the stacking fault energy (SFE).
Generally, materials with a low SFE prefer TRIP; a moderate SFE results in deformation via TWIP, while alloys with a high SFE deform via dislocation slip~\cite{Pierce2015-ja}.
These material characteristics can be influenced by chemical composition. 
Hence, alloying is a practical design tool for tuning the actual deformation mechanism~\cite{Zhao2021-an}.

Useful insights into the mechanics of the phase transformations can be obtained from atomistic simulations.
DFT revealed that the martensitic $\alpha''$ phase prefers to nucleate and grow at the $\{332\}\langle11\bar3\rangle$ twin boundary~\cite{Chen2019-ro} and therefore dominates the deformation mechanism in pure Ti.
Later,~\citet{Zhang2023-yz} argued that $\alpha''$ is unlikely to be directly formed at the twin boundary and is likely to be remnants of the pre-existing $\alpha''$.
Analogously, the $\omega$ phase has been proposed to promote the $\{112\}\langle11\bar1\rangle$ twinning in the Ti-Mo systems~\cite{Chen2021-am}.

\citet{Wang2023-sd} employed a custom-made molecular dynamics (MD) potential to study $\beta\leftrightarrow\alpha$ martensitic transformation- and dislocation-behavior in the Ti80 alloy (Ti–6Al–3Nb–2Zr–1Mo), confirming the bcc-hcp Burgers orientation relationship (OR).
$\beta\leftrightarrow\omega$ transformation barriers for several $\beta$-stabilizers in Ti-$X$ binary alloys were reported by~\citet{Salloom2021-qj}, who linked them with trends in valence electron concentration (VEC) and Young's modulus.
\citet{Chen2024-fr} studied the $\beta$-Ti-Nb-Mo system using a combination of MD and DFT to reveal that an increase in the VEC leads to an avoidance of the $\omega$ phase formation at the twin-boundary, thereby lowering the shear stress needed for twin growth. 
\citet{Wang2024-sd} concluded from their MD simulations that increasing Nb content in the $\beta$-Ti-Nb system leads to a transition from twinning to dislocation slip.
Similarly, \citet{Chen2019-bf} showed, using DFT and experiments, that structural evolution of the $\omega$ phase in the Ti-20at.\%Mo system promotes the transition from twinning to the dislocation slip mechanism.

The focus of this work is to apply \emph{ab initio} calculations to characterize and better understand the energy landscapes accompanying TRIP and TWIP transformations and to provide fundamental data for a series of nine Ti-$X$ ($X=$Al, Si, V, Cr, Fe, Cu, Nb, Mo, Sn) binary systems, which can be used in subsequent alloy development.

\section{Methods}
\label{sec:methods}

\subsection{Models of deformation mechanisms}
\label{sec:defomrmation_models}
In total, four transformation models based on the overview in the introduction are considered in the present work: two for the TRIP deformation ($\beta\leftrightarrow\alpha$ and $\beta\leftrightarrow\omega$) and two for twinning. 
Due to the structural similarity between the $\alpha$ and $\alpha'$ phases, both are simulated using hcp structures, and no distinction between them is made in the present work.

\begin{figure}[h]
    \centering
    \includegraphics[width=0.8\textwidth]{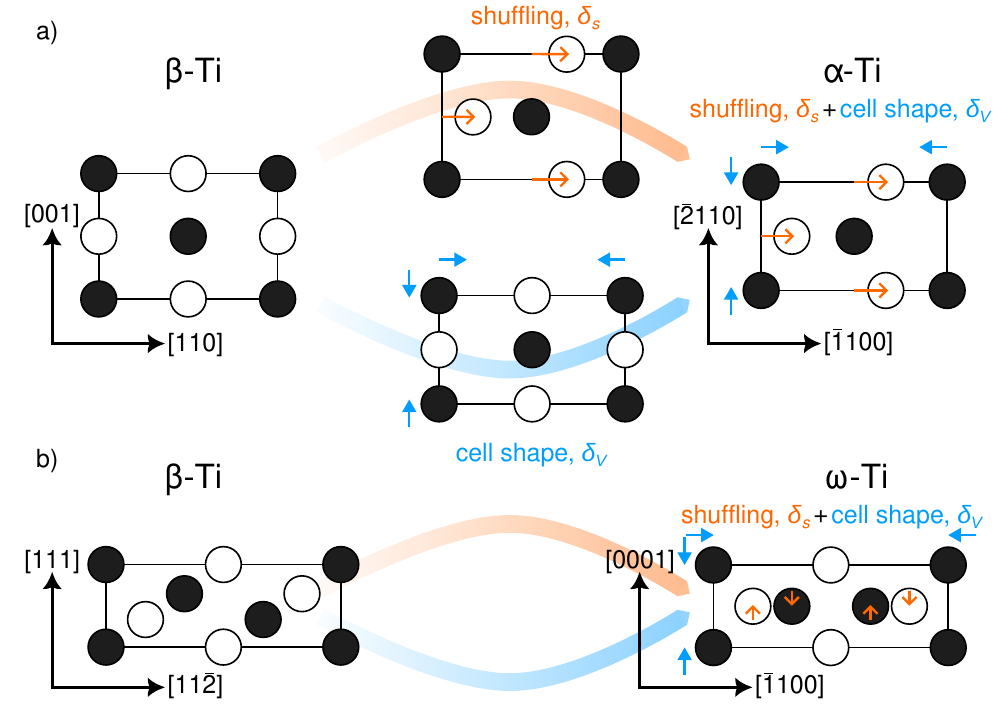}
    \caption{Atomistic mechanisms of the a) $\beta\leftrightarrow\alpha$ and b) $\beta\leftrightarrow\omega$ TRIP.
    In both cases, the transformation can be decoupled to atomic shuffling, $\delta_s$ (orange), and cell shape, $\delta_V$ (blue).
    In the $\beta\leftrightarrow\omega$ transformation, the volumetric changes are minute and hence not explicitly shown.
    }
    \label{fig:TRIP}
\end{figure}

The $\beta\leftrightarrow\alpha$ transformation is based on the Burger's orientation relationship $(110)_{\text{bcc}}\|(0001)_{\text{hcp}}$, which has been detailed in Ref.~\cite{Abdoshahi2022-lg}.
Thereby, the configurational space is described with two independent parameters: shuffling of the atoms, $\delta_s$, and volume change\add{/cell shape}, $\delta_V$.
These are chosen so that $\delta_s=0, \delta_V=0$ represents an ideal bcc phase, while $\delta_s=1, \delta_V=1$ corresponds to the hcp phase.
\add{The cells in between, e.g. for $0<\delta V<1$, are constructed by linearly interpolating the lattice parameters between those of the bcc and hcp structures, thereby simultaneously changing cell shape and volume.}
This description is achieved by using a 4-atomic unit cell shown in Fig.~\ref{fig:TRIP}a.

A 6-atomic unit cell shown in Fig.~\ref{fig:TRIP}b, composed of three $(112)$ bcc planes, allows to define the $\beta\leftrightarrow\omega$ structural connection: in-plane shuffling of two planes is described by $\delta_s$, while the volume expansion \add{and cell shape changes are}\del{is} controlled by $\delta_V$. 
\add{Note that the cell shape and volume are coupled in the present treatment and both are uniquely defined by linearly interpolating the lattice parameters between those of the bcc and the $\omega$ phases.}

For the $\{332\}/\langle113\rangle$ twinning, the direct $\beta\leftrightarrow\beta_T$ transformation was implemented according to Ref.~\cite{Kwasniak2022-xv}.
Specifically, a 22-atomic unit cell with lattice vectors $\vec a=a_\beta/2[\bar3\bar11]_\beta$, $\vec b=a_\beta[0\bar11]_\beta$ and $\vec c=a_\beta[2\bar33]_\beta$ ($a_\beta$ is the bcc lattice parameter) was constructed; in this setting, the twinning $(2\bar33)$ planes are perpendicular to the $\vec c$ direction.
All atoms are then shifted by $\zeta\vec u$, where $\zeta$ is the transformation coordinate ($\zeta=0$ corresponds to $\beta$, $\zeta=1$ to $\beta_T$).
The displacement vector $\vec u$ is given by:
\begin{equation}
    \vec u = \left[z\frac{a}2, 0, -\sgn(y-0.25) \frac{c}{22}\right]\ ,
    \label{eq:displ_u}
\end{equation}
where $a$ and $c$ are lengths of lattice vectors $\vec a$ and $\vec c$, $y$ and $z$ are fractional coordinates in the model unit cell (see \SI for more details).
These shifts are realized together with the overall cell shape change due to shear caused by tilting the lattice vector $\vec c$ to $\vec c+\zeta\vec a$.

The atomistic shuffling of the $\{112\}\langle11\bar1\rangle$ twinning \change{are}{is} inspired by the mechanism described in Ref.~\cite{Tobe2014-gh}.
The initial structure was a 6-atomic unit cell with lattice vectors $\vec {a'} = a_\beta/2[\bar1\bar11]$, $\vec {b'} = a_\beta[1\bar10]$ and $\vec {c'}=a_\beta[112]$.
The shear caused by tilting $\vec {c'}$ to $\vec {c'}+\zeta \vec {a'}$ is accompanied by additional shuffling on every second $(112)$ plane in the $\vec {a'}$ direction:
\begin{equation}
    \vec u_{2} = (-\zeta a', 0, 0)\ ,\quad \vec u_{4, 6} = (\zeta a', 0, 0)\ , \quad \vec u_{1,3,5} = (0, 0, 0)\ ,
\end{equation}
where the index represents the $(112)$ planes.

The structural models are detailed in \SI.
The implementation of those models into the Atoms Python object of Atomic Simulation Environment (ASE)~\cite{Larsen2017-dx} is available through an open-access git repository~\cite{gitrepo}.

\subsection{Computational details}
\label{sec:DFT}
Density Functional Theory (DFT) calculations were performed utilizing the Vienna Ab-initio Simulation Package (VASP)~\cite{Kresse1996-gt, Kresse1996-tg}.
Projector augmented plane-wave (PAW) pseudo-potentials~\cite{Kresse1999-if} were employed for the electron-ion interactions and the electronic exchange and correlation effects were treated at the level of the generalized gradient approximation (GGA)~\cite{Perdew1996-vd}. 
The reciprocal space was sampled with a Gamma-centered Monkhorst-Pack scheme with evenly spaced $k$-points at a maximum spacing of $0.041\uu{\AA}^{-1}$.
A plane-wave cutoff energy was set to $500\uu{eV}$.
A stopping criterion of $10^{-4}\uu{eV}$ per supercell was used for the charge density convergence.
These settings guarantee the total energy accuracy of $\approx 1\uu{meV/at.}$

To deal with structural disorder, Special Quasi-random Structures (SQS) \cite{Wei1990-zt} were generated using the \texttt{sqsgenerator} code~\cite{Gehringer2023-vu}.
Differently-sized supercells aiming at similar compositions (to allow the comparison of the alloying trends) were constructed, considering the computational resources needed and the necessary geometrical constraints.
In the TRIP calculations, we employed $3\times4\times3$ supercells (144 atoms, out of which 20 sites were occupied by species other than Ti, i.e., $x_X=0.139$) for the $\beta\leftrightarrow\alpha$ potential energy surface (PES).
Smaller $3\times2\times2$ cells (36 atoms, 5 alloyed sites, i.e., $x_X=0.139$) were used for the $\beta\leftrightarrow\omega$ PES calculations.
The $\{332\}\langle113\rangle$ twinning was modeled using $1\times2\times1$ supercells (44 atoms, 6 alloyed sites, i.e., $x_X=0.136$), and finally, $3\times2\times2$ supercells (36 atoms, 5 alloyed species, i.e. $x_X=0.139$) were utilized for the $\{112\}\langle11\bar1\rangle$ twinning mechanism.

Compositionally-dependent structural parameters were obtained by fully relaxing the alloyed models (cell volume and shape as well as ionic positions) and are reported in \SI.
The unrelaxed structures were subsequently scaled to fit the respective equilibrium volume, and the TRIP and TWIP calculations were performed without relaxing the atomic positions (while considering the equilibrium volumes).

The complete DFT data are available as a NOMAD repository dataset~\cite{nomad_dataset}.

\section{Results and Discussion}
\label{sec:res_and_dis}

\subsection{Energetics of the $\alpha\leftrightarrow\beta(\leftrightarrow\omega)$ transformation}

\begin{figure}[h]
    \centering
    \parbox{0.49\linewidth}{(a)}
    \parbox{0.49\linewidth}{(b)}
    
    \includegraphics[width=0.49\linewidth]{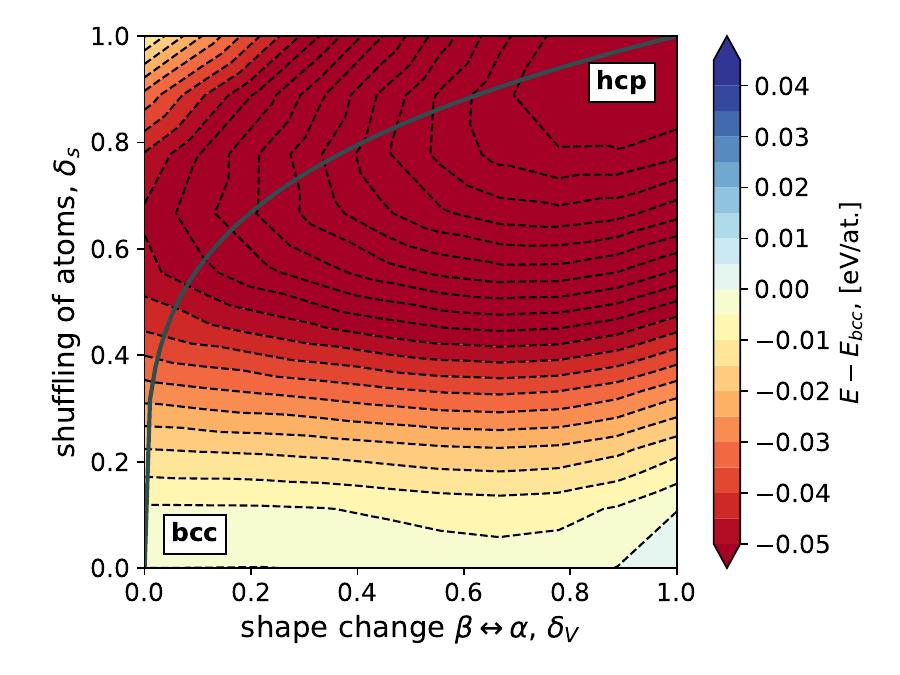}
    \includegraphics[width=0.49\linewidth]{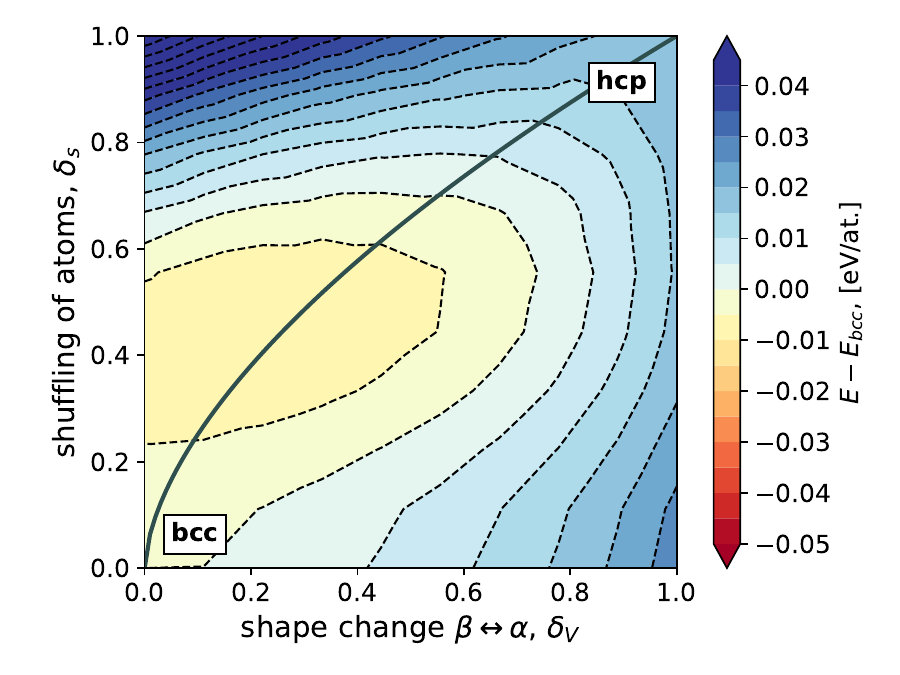}    
    \caption{Examples of TRIP-PES: (a) pure Ti, (b) Ti-14at.\%Mo solid solution (\ce{Ti_{0.86}Mo_{0.14}}).
    The dark green lines are guides for the eye to depict the steepest gradient descent connection of the bcc-$\beta$ ($\delta_V=0, \delta_s=0$) and hcp-$\alpha$ ($\delta_V=1, \delta_s=1$) corners.
    }
    \label{fig:TRIP-PES}
\end{figure}

Representative examples of the transformation PES are shown in Fig.~\ref{fig:TRIP-PES}. 
The Ti hcp-$\alpha$ structure ($\delta V=1$, $\delta s=1$) is more stable by $109\uu{meV/at.}$ than the bcc-$\beta$ phase ($\delta V=0$, $\delta s=0$) (Fig.~\ref{fig:TRIP-PES}a).
Importantly, there is no energy barrier along the path connecting the two structures with the Burger's orientation relationship.
This means that $\beta$-Ti spontaneously transforms to the $\alpha$-Ti, in correspondence with its instability at low temperatures.
Figure~\ref{fig:TRIP-PES}b shows the same PES for the case of \ce{Ti_{0.86}Mo_{0.14}}.
Firstly, Mo stabilizes the $\beta$ phase w.r.t. the $\alpha$ phase: $\Delta E_{\beta-\alpha}=-18\uu{meV/at.}$
Secondly, the global energy minimum is neither the bcc-$\beta$ nor the hcp-$\alpha$ phase, but instead some intermediate structure in between (see later).

\begin{figure}[h]
    \centering
    \parbox{0.49\linewidth}{(a)}
    \parbox{0.49\linewidth}{(b)}

    \includegraphics[width=0.49\linewidth]{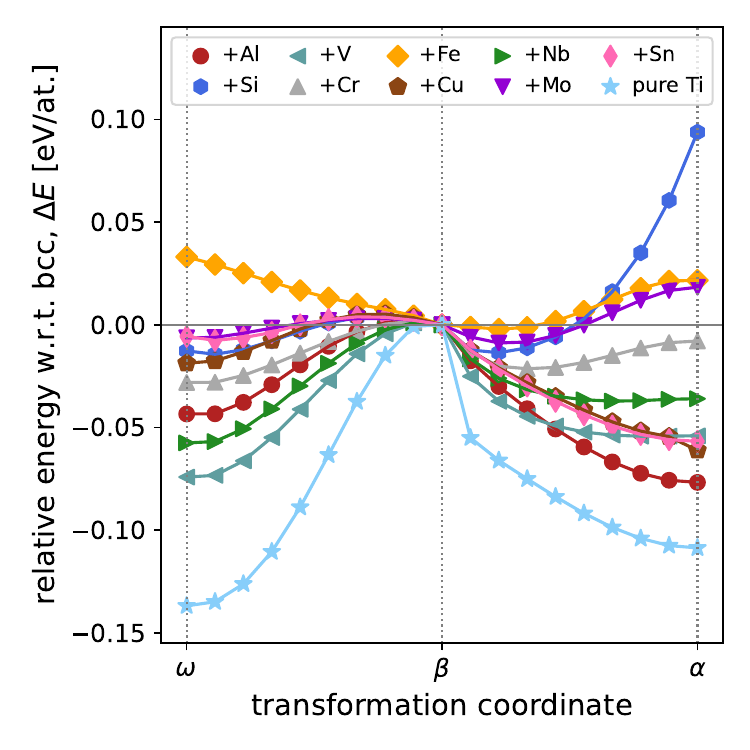}
    \includegraphics[width=0.49\linewidth]{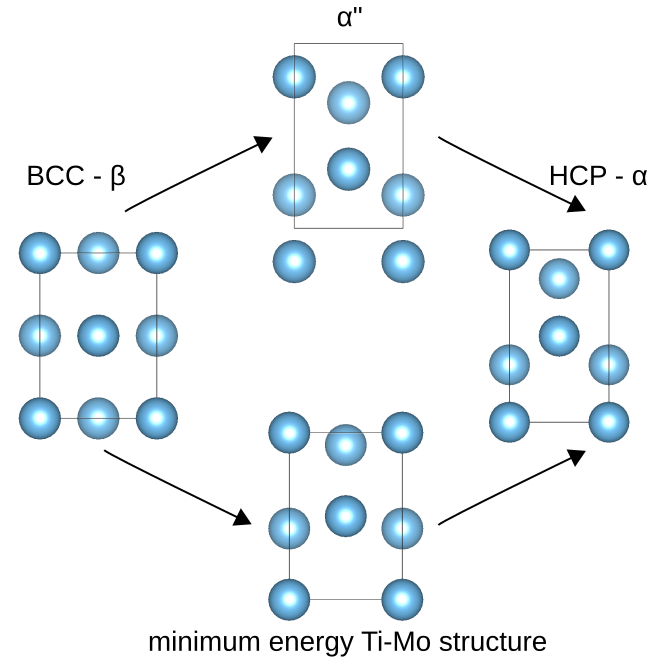}
    \caption{(a) $\alpha\leftrightarrow\beta(\leftrightarrow\omega)$ energetics.
    (b) Martensitic $\alpha''$ and minimum energy structure along $\beta\leftrightarrow\alpha$ path.}
    \label{fig:omega-beta-alpha_ene}
\end{figure}

The energies along the transformation path were extracted from individual PES (given in \SI for all here investigated species) by following the steepest gradient descent path, as visualized by the dark green lines in Fig.~\ref{fig:TRIP-PES}.
The resulting barriers are shown in the right section of Fig.~\ref{fig:omega-beta-alpha_ene}a (transformation coordinate between $\beta$ and $\alpha$).
Namely, 13.8\,at.\% of Al, V, Sn, Cu, and Nb decrease the energy preference for the $\alpha$ phase w.r.t. the $\beta$ phase, while the spontaneous character of the transformation remains.
In the case of Cr, the $\beta$ phase is only slightly less favorable than the $\alpha$ phase, whereas for Mo, Fe, and Si, it becomes actually the preferred configuration.
Interestingly, in all these four cases, the most stable configuration appeared somewhere in between the $\beta$ and $\alpha$ phases.

In Fig.~\ref{fig:omega-beta-alpha_ene}b, we visualize the projections of the structure corresponding to the energy minimum for the Ti-Mo system, along with the corresponding projections of the bcc-$\beta$ (110) and hcp-$\alpha$ (0001) planes.
Literature reports on an $\alpha''$ phase, an orthorhombic intermediate phase between bcc and hcp~\cite{Zheng2022-iu}.
The corresponding projection is also shown in Fig.~\ref{fig:omega-beta-alpha_ene}b.
Our minimum-energy structure resembles the $\alpha''$ phase, although, for the Ti-Mo system, it appears closer to the $\beta$ phase shuffling-wise than the $\alpha''$ phase.
This could also be related to the amount of Mo, which is a strong $\beta$ stabilizer~\cite{Huang2016-gf}. 
\citet{Raabe2007-lz} reported using DFT, that the $\beta$ is thermodynamically stable for about $25\uu{at.\%}$, which further reduces to $14\uu{at.\%}$ at $1154\uu{K}$.
It is, therefore, reasonable to expect a perfect (stable) bcc cell, hence the disappearance of the $\alpha''$ minimum, for concentrations only slightly higher than $14\uu{at.\%}$ Mo.
In contrast, the Cr minimum along the bcc-hcp transformation is shifted more towards the $\alpha$ phase and, hence, structurally closer to the $\alpha''$ phase.
Overall, it can be concluded that the addition of Mo and Cr not only stabilizes the $\beta$ phase but also promotes the $\alpha''$ phase, which is in agreement with experimental reports \cite{Barriobero-Vila2015-jz}.

The stable structure of Ti at $0\uu{K}$ is the $\omega$ phase.
Thus, the energetics of the $\beta\leftrightarrow\omega$ transformation was also investigated in a similar manner as for $\beta\leftrightarrow\alpha$.
In this case, the potential energy surface exhibits almost no dependence on the \add{volume/}cell-shape parameter $\delta V$, and all energy differences along the PES are almost exclusively related to shuffling, $\delta s$.
The individual PES are provided in \SI.
It was found that all investigated elements decrease the energy difference between the $\beta$ and the $\omega$ phase.
Thereby, all three phases, $\alpha$, $\beta$, and $\omega$, become energetically more alike upon alloying Ti.
For V and Nb, the $\omega$ phase remains the most stable (as is the case for pure Ti), consistent with previous calculations~\cite{Salloom2021-qj}.
This is also the case for Cr, which is closely followed by the $\alpha''$ phase.
For Al, Cu, and Sn, the hcp-$\alpha$ phase is the most stable.
Finally, the most stable phase in the case of the Ti-Mo system is the $\alpha''$.
Importantly, neither of the investigated systems (and compositions) yields the bcc-$\beta$ phase as the most stable configuration as a consequence of the $0\uu{K}$ calculation conditions.

\subsection{Twinning barriers}
\label{sec:twinning}

\begin{figure}[h]
    \centering
    \parbox{0.49\linewidth}{(a)}
    \parbox{0.49\linewidth}{(b)}
    
    \includegraphics[width=0.49\linewidth]{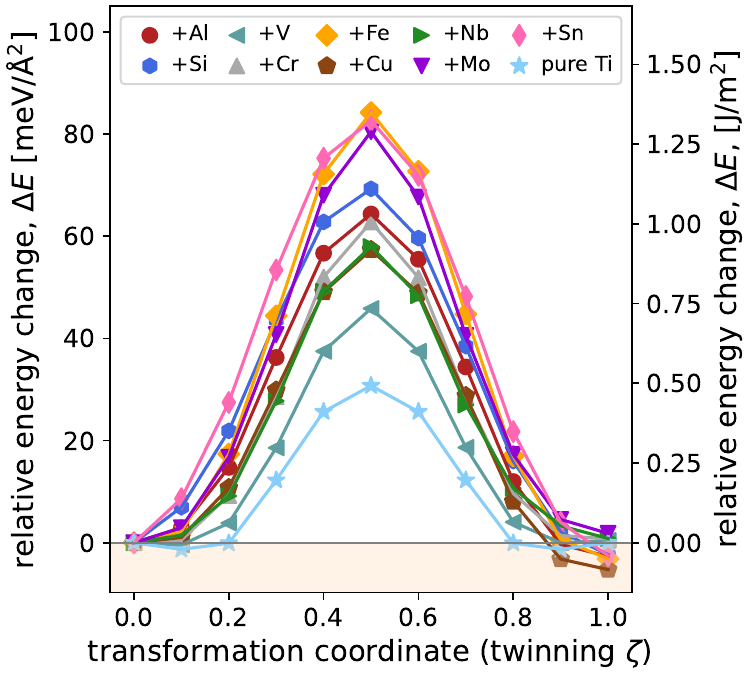}
    \includegraphics[width=0.49\linewidth]{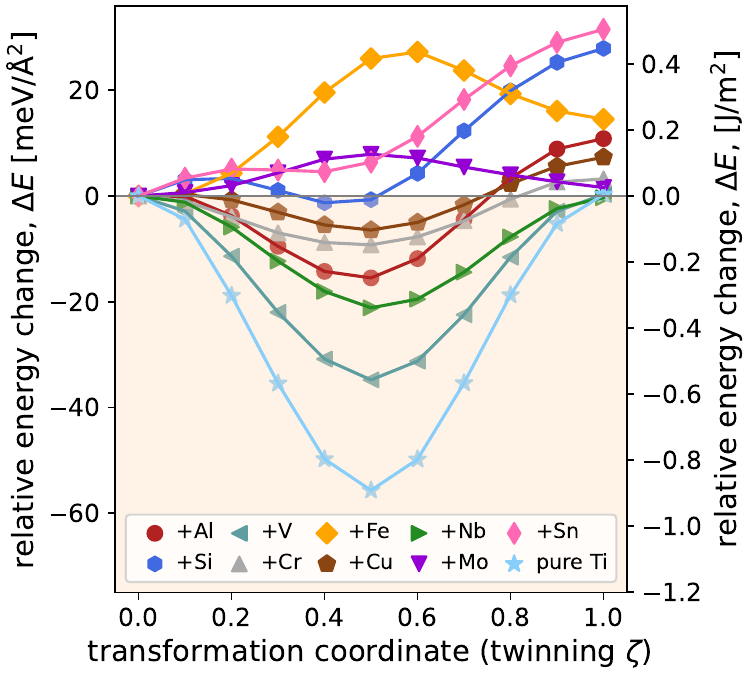}
    \caption{TWIP barriers for (a) $\{112\}\langle11\bar1\rangle$ and (b) $\{332\}\langle11\bar3\rangle$ twinning mechanisms.}
    \label{fig:TWIP_barriers}    
\end{figure}

Considering the two previously suggested twinning mechanisms as relevant for bcc Ti.
For the $\{112\}/\langle11\bar1\rangle$ system (Fig.~\ref{fig:TWIP_barriers}a), the lowest barrier of $31\uu{meV/\AA}^2$ is obtained for pure Ti.
Alloying contents of $\approx 14\uu{at.\%}$ increase the twinning barrier for all species investigated here.
The smallest increase of $41\uu{meV/\AA}^2$ is predicted for V while Al, Si, Fe, Mo, and Sn more than double the barrier w.r.t. pure Ti.

\begin{figure}[h]
    \centering
    \parbox{0.48\linewidth}{(a)}\hfill
    \parbox{0.48\linewidth}{(b)}

    \raisebox{5mm}{\includegraphics[width=0.48\linewidth]{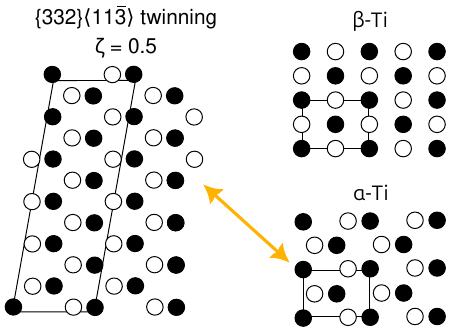}}\hfill
    \includegraphics[width=0.48\linewidth]{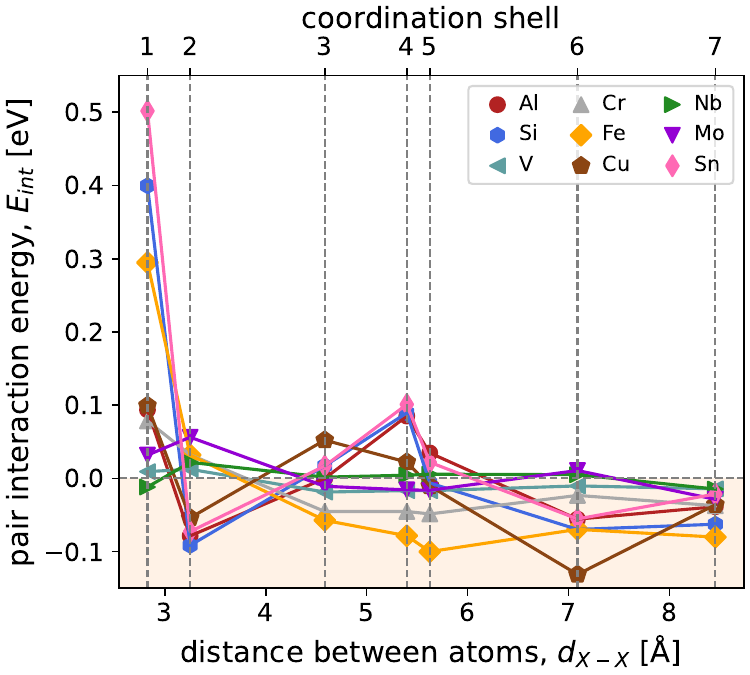}
    
    \caption{
    (a) Origin of the negative barrier. 
    The atomic arrangement at $\zeta=0.5$ resembles that of the $\alpha$-Ti. 
    (b) Pair interaction energies for different species as functions of their distance expressed in Angr\"om (lower axis) and coordination shell (upper axis).
    }
    \label{fig:Eint}
\end{figure}

Considering the $\{332\}\langle11\bar3\rangle$ system, the results are more complex and do not correlate directly with the commonly expected behavior.
Pure Ti exhibits a deep minimum of $-56\uu{meV/\AA}^2$ halfway along the transformation path. 
The atomic arrangement corresponding to this point, shown in Fig.~\ref{fig:Eint}a, closely resembles the atomic structure of the hcp-$\alpha$ Ti.
Keeping in mind the PES of pure Ti (Fig.~\ref{fig:TRIP-PES}a) clearly preferring hcp over bcc, this minimum becomes a natural consequence of this phase preference.
All other alloying elements make this minimum shallower, again, in agreement with the fact that alloying decreases the energy difference between the $\beta$ and the $\alpha$ phases (Fig.~\ref{fig:omega-beta-alpha_ene}a).
Finally (and again in agreement with the $\alpha\leftrightarrow\beta$ PES), Mo and Fe yield a small positive energy barrier.

It is worth noting that Si, Sn (but also, e.g., Mo to a lesser degree) exhibit a non-symmetrical shape of their $\{332\}\langle11\bar3\rangle$ energy landscape.
To elucidate on this, we investigated the solute pair-interaction energies in the bulk bcc Ti matrix (Fig.~\ref{fig:Eint}b).
While for a majority of time these interactions are negligible, a strong repulsion between first nearest neighbors of Sn-Sn, Si-Si and Fe-Fe exists.
The structural SQS models have optimized short-range order parameters; this is the initial cell for $\zeta=0$.
Counting the number of X-X nearest neighbor pairs (X is the alloying element), we get 3 pairs per simulation cell. 
The maximum number of these pairs connected to a single X site is 2 (i.e. there exists at least 1 X atom, for which 2 out of 8 nearest neighbors are also X atoms).
Performing the same analysis for the transformed cell ($\zeta=1$), we obtain 5 X-X pairs, with 1 X site having 4 X atoms as nearest neighbors.
Therefore, the $\{332\}\langle11\bar3\rangle$ twin contains more X-X nearest-neighbor pairs, which increase the total energy in the case of Si, Sn, and Fe due to their repulsive interaction.
This justifies the non-symmetrical shape of the Si, Sn, and Fe curves in Fig.~\ref{fig:TWIP_barriers}b; for the other species, the interaction energies are significantly smaller, and hence, the energy difference between the structure at $\zeta=0$ and $\zeta=1$ are only minute.

Inspecting the scatter of energies for $\zeta=1$ in the $\{112\}/\langle11\bar1\rangle$ case, no such large differences are observed. 
This is because the number of X-X bonds remains similar for $\zeta=0$ (4 X-X nearest neighbor pairs, maximum of 2 X atoms bonded to one X) and $\zeta=1$ (3 X-X nearest neighbor pairs, maximum of 3 X atoms bonded to a single X).

It is noted that this energy scatter is a consequence of the used relatively small simulation cells and can be overcome by using sufficiently large cells or by performing a statistical approach by averaging many small cells.
Nevertheless, the alloying-related trends, as well as the fact that the twinning barrier is negative or positive (w.r.t. to bcc structure), remain qualitatively the same.

\begin{figure}[h]
    \centering
    \parbox{0.48\linewidth}{(a)}\hfill
    \parbox{0.48\linewidth}{(b)}

    \includegraphics[width=0.48\linewidth]{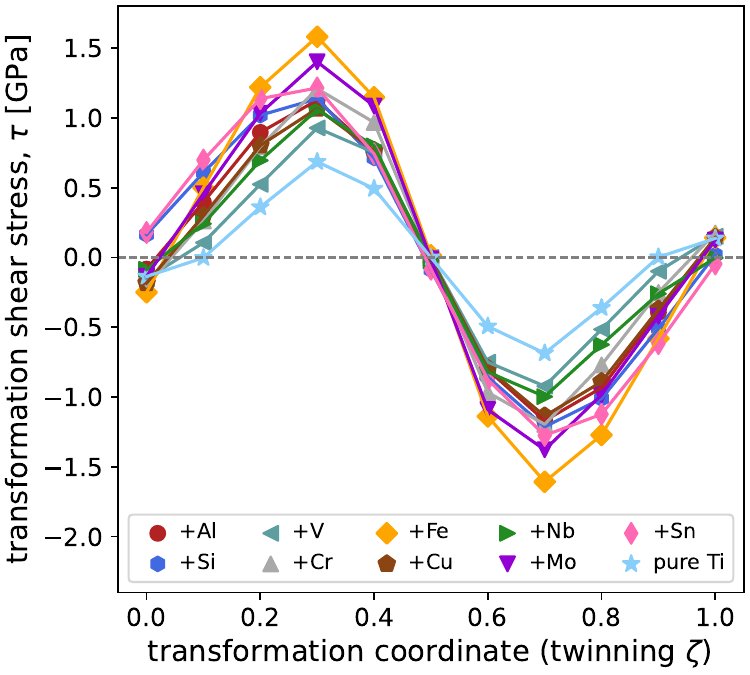}
    \includegraphics[width=0.48\linewidth]{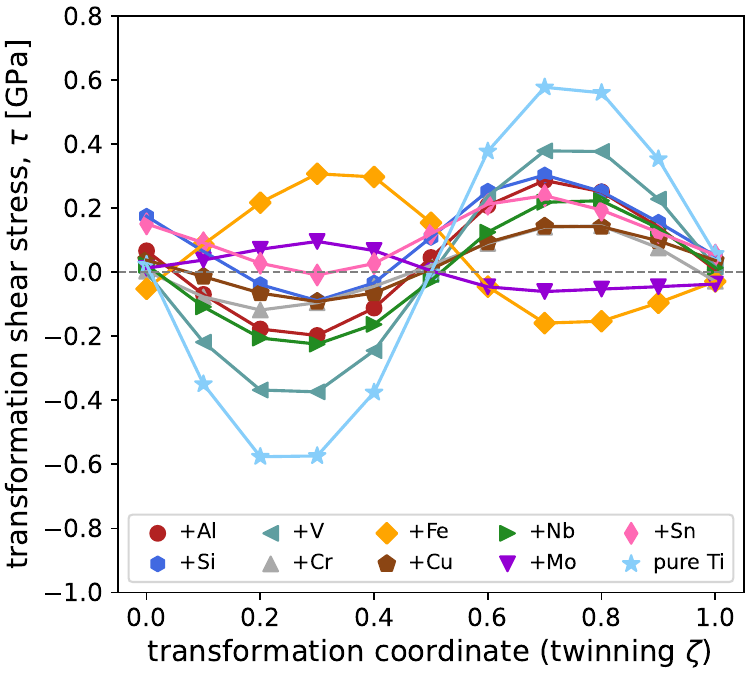}
    
    \caption{
    Transformation shear stress calculated from the energy profiles in Fig.~\ref{fig:TWIP_barriers} using Eq.~\eqref{eq:tau} for (a) $\{11\bar2\}\langle11\bar1\rangle$ and (b) $\{33\bar2\}\langle11\bar3\rangle$ twinning.
    }
    \label{fig:tau}
\end{figure}

Having the energy landscape for the two twinning mechanisms (Fig.~\ref{fig:TWIP_barriers}), it is possible to predict the alloying impact on the theoretical shear strength.
This is done by calculating the transformation shear stress as
\begin{equation}    
    \tau(\zeta) = \frac1{2^{3/2}}\frac{1}{V}\frac{d E_{\mathrm{tot}}(\zeta)}{d\zeta}
    \label{eq:tau}
\end{equation}
where the numerical prefactor is related to the aspect ratio of the used cells (and is identical for both twinning mechanisms), $V$ is the volume of each cell (remains constant during the deformation due to the model setup), and $E_{\mathrm{tot}}$ is the total energy as a function of the transformation coordinate $\zeta$.
The resulting plots are shown in Fig.~\ref{fig:tau}.
For the stable shearing scenario of $\{112\}/\langle11\bar1\rangle$ twinning, alloying increases the theoretical shear stress from $\approx 0.7\uu{GPa}$ for the pristine Ti up to $\approx 1.6\uu{GPa}$ caused by adding $\approx 14\uu{at.\%}$ of Fe.
In the case of unstable $\{332\}/\langle11\bar3\rangle$ twinning, alloying generally decreases the instability, thereby reducing the maximum shear stress driving the $\beta\to\alpha_{\zeta=0.5}$ transformation from $\approx -0.6\uu{GPa}$ to below $-0.1\uu{GPa}$ for Cu, up to positive barriers of $\approx 0.3\uu{GPa}$ and hence stabilization of the $\beta$ twins via alloying Fe.
These values are significantly larger than the real transformation stresses measured experimentally~\cite{Gao2018-to}.
This is because our models implement homogeneous strain over the sample and thereby correspond to the upper theoretical limits of the shear strength.

\subsection{Alloying impact on stacking fault- and twin boundary-energies}

To further assess the impact of alloying on the stacking fault and twin boundary energies, models with decorated monolayers of alloying species will be discussed in this section. 
The perfect bcc structure consists of a repetitive stacking of six $\{11\bar2\}$ planes denoted as $A$, $B$, $C$, $D$, $E$, and $F$.
Applying a shift of $\frac{a}{6}[111]$ to a half-crystal staring with plane $F$ and above, the stacking order becomes:
$$
\begin{array}{llllllllllllll}
    \dots & A & B & C & D & E & F & A & B & C & D & E & F & \dots\\
          &   &   &   &   &   &  & \downarrow & \downarrow & \downarrow & \downarrow & \downarrow & \downarrow & \\
    \dots & A & B & C & D & E & \underline{F} & \underline{E} & F & A & B & C & D & \dots \\
\end{array}
$$
This way, a stacking fault (SF) is created (visualized above by the underlined planes). 
Noteworthily, this SF can also be seen as a twin nucleus: applying another $\frac{a}{6}[111]$ shift to the planes $F$ and above results in enlarging the twin from two to three planes and so on.
We, therefore, denote this configuration as a \textit{nanotwin}.

Since the model consists of 12 layers, after six repetitions, we arrive at a model with two equally sized twin grains consisting of 6 layers each, i.e. on two maximally separated twin boundaries.
$$
\begin{array}{llllllllllllll}
    \dots & A & B & C & D & E & F & A & B & C & D & E & F & \dots\\
          &   &   &   &   &   &  & \downarrow & \downarrow & \downarrow & \downarrow & \downarrow & \downarrow & \\
          &   &   &   &   &   &  &  & \downarrow & \downarrow & \downarrow & \downarrow & \downarrow & \\
          &   &   &   &   &   &  &  &  & \downarrow & \downarrow & \downarrow & \downarrow & \\
          &   &   &   &   &   &  &  &  &  & \downarrow & \downarrow & \downarrow & \\
          &   &   &   &   &   &  &  &  &  &  & \downarrow & \downarrow & \\
          &   &   &   &   &   &  &  &  &  &  &  & \downarrow & \\
    \dots & A & B & C & D & E & F & \underline{E} & \underline{D} & \underline{C} & \underline{B} & \underline{A} & \underline{F} & \dots \\
\end{array}
$$
where the underlined planes correspond to the twin grain.
The mirror planes in this example are the $F/\underline{F}$ planes.
We call this configuration as \textit{separated twin GBs}.

The twin grain boundary (GB) energy is calculated as:
\begin{equation}
    \gamma_\text{twin}^\text{Ti} = \frac1{2A}(E(\mbox{2 twin GBs}) - E(\mbox{perfect crystal})) \approx 120\uu{mJ/m}^2\ ,
\end{equation}
where $A$ is the area of the $\{11\bar2\}$ plane in the unit cell. 
``slab with 2 twin GB'' are, in our case, either the nanotwin or the separated twin GBs geometry.
Both of them yield $\gamma$ within $1\uu{mJ/m}^2$ for pure Ti. 
The values agree well with the previously calculated DFT value of $100\uu{mJ/m}^2$~\cite{Zhang2023-yz}.

To estimate the alloying impact on the twin GB energy, we populated one plane adjacent (e.g. plane $\underline{E}$) to the GB fully with a species $X$ and evaluated the energy of the decorated twin boundary as:
\begin{equation}
    \gamma_\text{twin}^X = \frac1A(E(\mbox{1 decorated and 1 pure GB}) - E(\mbox{1 decorated plane in perfect crystal}))-\gamma_\text{twin}^\text{Ti}\ .
\end{equation}
The last term accounts for the second twin GB, which in the model remains undecorated.

\begin{figure}[h]
    \centering
    \parbox{0.49\linewidth}{(a)}\hfill
    \parbox{0.49\linewidth}{(b)}
    
    \includegraphics[width=0.49\linewidth]{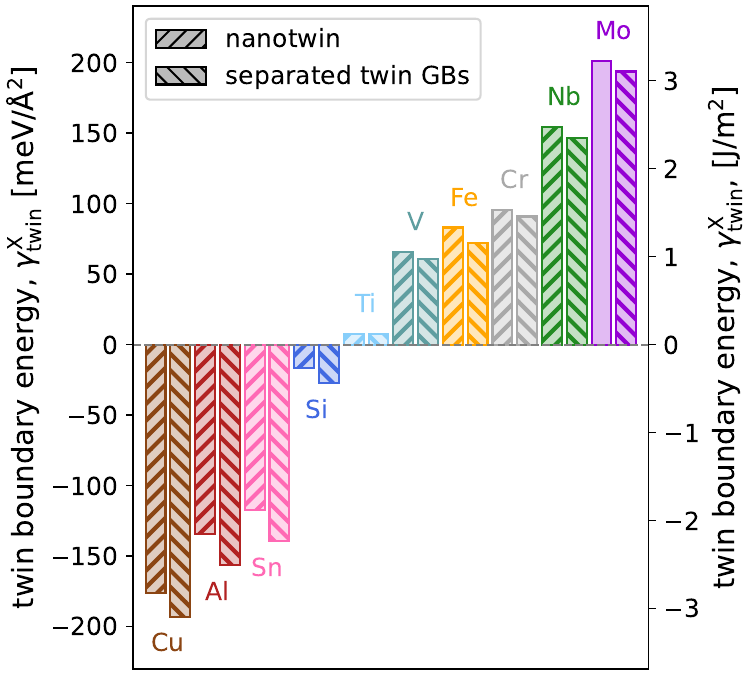}\hfill
    \includegraphics[width=0.49\linewidth]{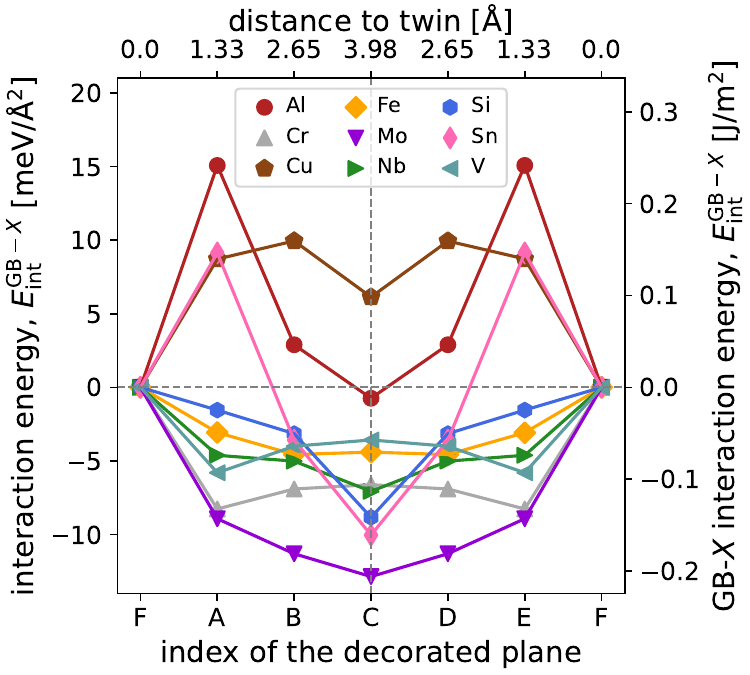}
    \caption{(a) Calculated twin boundary energies of a single decorated layer located at the twin boundary for nanotwin (left bars) and separated twin boundary (right bars) configuration. (b) Interaction energies between twin boundary and the plane decorated by alloying species. Positive (negative) values correspond to repulsion (attraction) of the twin boundary and the alloying element.
    }
    \label{fig:SFE}
\end{figure}

The results shown in Fig.~\ref{fig:SFE}a firstly suggest no qualitative difference between nanotwin and secondly separate twin GBs scenarios and separate the alloying elements into two groups.
Cu, Al, Sn, and to a much smaller magnitude also Si, yield negative values for $\gamma_\text{twin}$. 
This means that twin GBs are energetically preferable, and hence, a thermodynamic driving force for twinning exists. 
On the contrary, V, Fe, Cr, Nb, and Mo (in this order) increase $\gamma_\text{twin}$ w.r.t. that of pure Ti, and, therefore, are expected to act against the twinning process.
Experimental works have shown that TWIP is encountered in alloying systems containing Sn, e.g., Ti-Cr-Sn~\cite{Lilensten2020-nz}.
It should be noted that the present calculations were done with a rather artificial configuration of having all alloying elements present as a monoatomic plane and thus should be interpreted only qualitatively.
Nevertheless, the trend is qualitatively in agreement with the trends reported for Ti-V, Ti-Nb, and Ti-Mo binary alloys calculated by~\citet{Zhang2023-yz} with a model having a homogeneous composition.

While in the above, it was considered that the decorated plane was directly adjacent to the twin GB, this does not have to be the case.
Figure~\ref{fig:SFE}b shows the relative $\gamma_\text{twin}^X$ changes depending on the position of the decorated plane from the GB: positions $F$ correspond to the scenario presented in Fig.~\ref{fig:SFE}a as ``separated twin GBs'', position $C$ correspond to the decorated plane being in the middle of one of the two (identical) twins.
Those results suggest that Mo, Nb, Si, and Fe anti-segregate away from the GB (will prefer to remain in bulk); Cr and V segregate to the plane next to the twin GB.
Sn and Al prefer the bulk (lowest energy at the plane $C$ farthest from the twin GBs) and exhibit a barrier at the adjacent plane next to the GB; if the GB gets decorated, the trapped atoms would hinder GB mobility.
Lastly, Cu exhibits a strong tendency to decorate the GB with a minor barrier w.r.t. the bulk position.

\section{Conclusion}
\label{sec:conlusion}

We have presented a comprehensive overview of transformation energy landscapes for $\alpha\leftrightarrow\beta\leftrightarrow\omega$ transformations in alloyed bcc Ti predicted by first-principles calculations.
All alloying elements, Al, Si, V, Cr, Fe, Cu, Nb, Mo, and Sn decrease the energy differences between the phases.
For V and Nb, the $\omega$ phase remains the most stable (as is the case for pure Ti), Al, Sn, and Cu yield the $\alpha$ phase as the most stable one.
Significant stabilization of the $\beta$ phase is predicted for Si, Cr, Fe, and Mo; however, in these cases, the martensitic $\alpha''$ phase emerges between the $\beta$ and $\alpha$ phases.

Further, twinning barriers for the two mechanisms were predicted. 
For the $\{112\}\langle11\bar1\rangle$ twinning, all alloying elements increase the barrier compared to pure Ti.
Since the intermediate state of the $\{332\}\langle11\bar3\rangle$ twinning is close to the hcp-$\alpha$ structure, all alloying species (including pure Ti) preferring the $\alpha$ phase over the $\beta$ phase yield negative barriers. 
A small positive barrier (hence stabilizing the $\beta$ phase) is predicted for Mo and Fe.

Finally, the alloying impact on stacking fault and twinning energies (on the $\{11\bar2\}$ planes) has been estimated.
Cu, Al, Sn, and Si yield negative twinning energies (i.e., thermodynamically preferable configurations), whereas V, Fe, Cr, Nb, and Mo yield significantly larger twinning energies than pure Ti, hence stabilizing the material against twinning.
Only Cu is predicted to segregate to the twin GB (and hence to be impactful). 
Interestingly, Al and Sn exhibited significant barriers between sitting directly in the GB and sitting in the adjacent plane, which may impact the mobility of the twin boundary (slowing down or even hindering the twinning transformation).

\backmatter

\bmhead{Supplementary information}

Supplementary information is provided as a separate pdf file.

\bmhead{Acknowledgements}

This research has been funded by the Austrian Federal Ministry for Climate Action, Environment, Energy, Mobility, Innovation and Technology (BMK) within the aeronautics funding program ``Take Off'', project ``T\textsuperscript{3}Design - TRIP and TWIP as Key Tools to Design Advanced AM Alloys for Aviation Structures''(grant agreement no. 903039) administered by FFG.
Calculations were performed using supercomputer resources provided by the Vienna Scientific Cluster (VSC).

\section*{Declarations}

\begin{itemize}
\item \textbf{Funding}: 
    This research has been funded by the Austrian Federal Ministry for Climate Action, Environment, Energy, Mobility, Innovation and Technology (BMK) within the aeronautics funding program ``Take Off'', project ``T\textsuperscript{3}Design - TRIP and TWIP as Key Tools to Design Advanced AM Alloys for Aviation Structures''(grant agreement no. 903039) administered by FFG
\item \textbf{Conflict of interest/Competing interests}: 
    We declare that there are no conflict of interest.
\item \textbf{Ethics approval and consent to participate}: Not applicable
\item \textbf{Consent for publication}: Not applicable
\item \textbf{Data availability}:
    The complete DFT dataset presented in this paper is shared as a NOMAD repository dataset: \url{http://dx.doi.org/10.17172/NOMAD/2025.04.14-2}.
\item \textbf{Materials availability}: Not applicable
\item \textbf{Code availability}: 
    Supporting routines used in pre-processing the structures for this work are shared as a Python package: \url{https://github.com/MUL-CMS/t3design}.
\item \textbf{Author contribution}:\\
    David Holec: Conceptualization, Methodology, Software, Validation, Resources, Writing - Original Draft, Writing - Review \& Editing, Visualization, Supervision, Project administration, Funding acquisition \\
    Johann Grillitsch: Methodology, Software, Validation, Formal analysis, Investigation, Data Curation, Writing - Review \& Editing, Visualization \\
    José L. Neves: Conceptualization, Validation, Resources, Writing - Review \& Editing \\
    David Obersteiner: Conceptualization, Validation, Investigation, Writing - Review \& Editing \\
    Thomas Klein: Conceptualization, Validation, Resources, Writing - Review \& Editing, Supervision, Project administration, Funding acquisition
\end{itemize}



\end{document}